\documentstyle[osa,manuscript,psfig]{revtex}  % DON'T CHANGE

\begin{document}                % INITIALIZE - DONT CHANGE
\title{ Bose-Einstein Correlations and Sonoluminescence}
\author{S. Trentalange$^*$ and S.U. Pandey$^{\dagger}$}
\address{$^*$Physics Department, University of California, Los Angeles, California 90024}
\address{$^{\dagger}$Physics Department, Ohio State University, Columbus, Ohio 43210 } 
\maketitle
\begin{abstract}               
Sonoluminescence may be studied in detail by intensity correlations among the emitted photons. As an example, we discuss an experiment to measure the size of the light-emitting region by the Hanbury Brown-Twiss effect.  We show that single bubble sonoluminescence is almost ideally suited for study by this method and that plausible values for the physical parameters are within easy experimental reach.  A sequence of two and higher order photon correlation experiments is outlined.
\end{abstract}
\section{Introduction} 
             
The field of sonoluminescence has undergone an explosion in recent years 
since the discovery of single bubble sonoluminescence (SBSL) [1-4].  In 
SBSL, a gas bubble in liquid is trapped at a velocity node of an 
acoustic field.  Under certain conditions, the bubble emits intense 
flashes of light ($> 10^{6}$ photons/flash ). The shape of the spectrum is the subject of much investigation; superficially it may be described as roughly that of a black-body at temperatures in excess of 20000 K.  Although the 
conditions for SBSL are now well-characterized for certain systems such 
as air bubbles in water, investigators have been denied a detailed look 
at the physical origins of this puzzling phenomenon due 
to the transient nature of the light flash.

For example, attempts to directly measure the duration of the flash have 
resulted only in an upper limit of 50-150 ps [5].  Likewise, while it 
is possible to make nanosecond scale measurements of the bubble 
size as a function of time by Mie scattering, this technique can 
resolve only the liquid-gas interface [6]; giving a value around 1 $\mu m$.  In any case, the suggested mechanism of shock wave separation and implosion would result in a much smaller radius for the physical process which converts the phonons to photons.

We propose the use of Bose-Einstein correlations to measure the spatial 
and temporal characteristics of the light-emitting region.  Bose-Einstein 
correlations have been used for many years in astronomy and 
nuclear physics to extend the accessible range of time and length scales.  
In these fields, the usefulness of this tool has been limited by 
the low brightness and non-reproducible nature of the sources, as 
well as final state interactions and unwanted correlations inherent in 
the particle production process [7].  These limitations are absent 
in the present case, and SBSL is ideally 
suited for application of this method.  

\section{Two-Photon Correlation Experiment.}

To illustrate these techniques, 
we now outline a specific two-photon experiment to measure the size of the 
light-emitting region in SBSL by the Hanbury Brown-Twiss (HBT) effect [7-10].  The two-photon correlation function 
$ C(k_{1},k_{2}) $ can be defined as

\begin{equation}
	C(k_{1},k_{2}) = P(k_{1},k_{2})/P(k_{1})P(k_{2})   .
\end{equation}
where $ P(k_{1},k_{2}) $ is the probability density of detecting photons of 4-momenta $k_{1}$ and $k_{2}$ in the same pulse and $P(k_{i})$ is the single photon probability density.  Because the total wavefunction for identical bosons is symmetric, the maximum value of this function will occur at $k_{1}=k_{2}$. When $k_{1}$ is not equal to $k_{2}$, the strength of the correlation will depend on the degree of overlap between the single particle wavefunctions of the two particles. This "exchange density" is determined by the size of the emitting source. Hence knowledge of the variation of $C(k_{1},k_{2})$ with the momentum difference carries information about the source distribution.  For completely uncorrelated photons, $C(k_{1},k_{2}) = 1$.  Thus, the HBT effect is maximal when the particles are emitted randomly. 

%in the case of SBSL this condition is guaranteed by the black-body nature of %the spectrum.

Correlation functions have been derived with varying degrees of rigor 
and generality [7,10]. Many of these concerns, however, have to do with 
mutual interactions of charged particles in the final state or residual 
coherence in the production process.  The absence of these effects in 
SBSL greatly simplifies the evaluation of Equation~1.  Assuming the photons are emitted with random relative phases and polarization states from a source with a normalized space-time density $\rho (r)$, the two-particle probability density can be written as an incoherent sum over the source amplitudes  $\Psi_{12}(k_{1},k_{2})$ for a pair of photons:

\begin{equation}
	P(k_{1},k_{2}) = \int |\Psi_{12}(k_{1},k_{2}) |^{2} \rho(r_{1}) \rho(r_{2}) d^{4}r_{1}d^{4}r_{2}    .
\end{equation}
while for the single particle densities
\begin{equation}
	P(k_{i}) = \int |\Psi_{i}(k_{i}) |^{2} \rho(r_{i})  d^{4}r_{i}    .
\end{equation}
Neuhauser [11] has derived an expression for the correlation function 
using symmetrized plane waves with unit directional vectors $\hat{ \bf{k}_{1}}$
and $\hat{ \bf{k}_{2}}$ as 
\begin{equation}
	C(k_{1},k_{2}) = 1 + \frac{1}{4} [ 1 + (\hat{ \bf{k}_{1}}\cdot \hat{ \bf{k}_{2}} )^{2} ] [ | \rho ( k_{1} + k_{2} ) |^{2} + | \rho ( k_{2} - k_{1} ) |^{2} ]    .
\end{equation}
where $ \rho (k) = \int e^{ i k r } \rho (r) d^{4}r $ is the Fourier 
transform of the source density.  The first factor in square brackets arises 
from the fact that there can be no interference between photons in different 
polarization states.  For the photon energies considered 
here $ \rho ( k_{1} + k_{2} ) \ll \rho ( k_{2} - k_{1} ) $ and 
can safely be neglected.  Thus, the correlation of colinear photons of 
equal energy is $C(k,k)= 3/2$.

To compare with experimental data, it is convenient to model the source.  Here we assume the photons to be emitted from a fireball with a stationary gaussian distribution of radius $ R $ and lifetime $ \tau $ [12]

\begin{equation}
	\rho (r,t) \propto exp[ - (r/R)^{2} - (t/ \tau )^{2} ]    .
\end{equation}
giving

\begin{equation}
	C(k_{1},k_{2}) = 1 + \frac{1}{4} [ 1 + \cos ^{2} ( \theta ) ]  exp[ - q^{2}R^{2} / 2 - \Delta E^{2}\tau ^{2} / 2 ]    .
\end{equation}
where $q = | k_{1} - k_{2} |$ is the relative momentum, $\Delta E = | E_{1} - E_{2} |$ is the energy difference of the 
photons and $\theta$ is the angle between them.  From the form 
of Equation~6, we can see that the width of the correlation function is inversely related to the source size $ R $ and the lifetime $\tau$.

To make the experimental situation more explicit, we begin with the data on SBSL reported in ref. [5] by Hiller, Putterman and Barber.  Here, a single air bubble trapped in water at room temperature emits $\sim$ 500,000 photons/flash at for a driving frequency of 27 kHz.  The measured spectral density has a high brightness in the visible region, which makes it possible to explore the most plausible "minimum bubble sizes" (10 nm - 10,000 nm) using commercially available photomultipliers, optical filters and other components.

The "minimum size" of the bubble, of course, refers strictly  to the diameter of the region where the light produced by as yet unknown mechanisms ceases to interact with the surrounding material and escapes.  Presumably, this occurs when the expanding material cools sufficiently to become transparent to the radiation.  A further step of interpretation must be made when one considers the refraction and polarization of the outgoing radiation through the mantle of material around it, as well as at the gas-liquid interface. However, these details had best be studied in accompaniment with direct measurements so as to be appropriate to the specific experimental conditions.

Estimates for the size and duration of the flash were obtained in a model investigated by Wu and Roberts [13] in which both the surface of the bubble and the enclosed gas were simulated. They integrated the Rayleigh-Plesset equation using a hard-core van der Waals equation of state for the gas.  From several reasonable scenarios presented one can expect typical "minimum bubble sizes" to range from 0.3-0.6 $\mu m$ and the duration of the event to be about 1 ps.  

If the light emitting region were about 0.1-1 $\mu m$ in radius, experiments 
to determine this number would be quite easy to perform.  
Figure~1 shows a schematic of an experimental setup. The momentum difference $q$
is solely a function of the energy and the separation angle of the two 
photons. For a fixed energy one can thus measure the correlation 
function by examining how Equation~1 varies with the separation angle. Using
Equation~6, the expected result of this measurement is seen in Figure~2
and is seen to extend over a 
large range of separation angles. We emphasize once more the 
transparent interpretation of HBT results using photons compared to 
charged particles. Final state strong interactions as well as Coulomb
repulsions  do not distort the signal when one uses photons. 
In addition, there is no ambiguity as to whether the photons one measures come from the source or arise from secondary processes such as $\pi^{0}$ decays. 

Realistic counting rates can be estimated for 
the geometry of Figure~1.
Two phototubes with high photoefficiency (5-10$\%$) at wavelengths approaching 200 nm could be masked with commercially available filters 
($\Delta \lambda \sim 10$ nm).  If the tubes are placed 20 cm from the source and further defined by 1 mm circular apertures, the real singles rate in each detector would be typically $\sim $ 0.1 photons/flash and the coincidence rate $\sim$ 0.01 photons/flash or about 200-300 counts/sec.  
For a source radius of 500 nm, the width of the correlation function  
corresponds to a detector separation of 20 mm.  Noise and random 
coincidences may be greatly reduced by triggering the phototube 
readout on the flash itself.  The choice of aperture size is somewhat arbitrary, and is governed by considerations of count rate and the scale of details one wishes to examine in the correlation function.

The high instantaneous brightness of the source ($> 10^{16}$ photons/sec) overcomes the objections to the coincidence method voiced in ref. [8].  However, different experimental modes may be contemplated if it becomes necessary to improve the signal to noise ratio in order to examine details of the correlation function.  In this case, the number of photons per flash striking each detector may be increased by using a larger aperture. If these values are stored in an ADC for each flash, then the system can be run as an intensity spectrometer, where correlations between the intensities  for each detector are computed for each flash and summed over many events.  This would be a digital version of the intensity correlator (linear multiplier) constructed by Hanbury Brown and Twiss [8].

Once the size of the bubble is known, it would be possible to directly measure the time duration of the flash in a conjugate experiment.  If the phototube positions are fixed within a region of large spatial correlation ($\ll$ 20 mm in the above example ) then the intensity correlation as a function of energy difference would give the lifetime.  Again, with commercially available filters, the photon energy difference could be stepped through a range of values which could probe flash lifetimes from 1 fs to 1 ps.

\section{Further Applications  }

Several extensions of the HBT experiment are immediately obvious.  First, the partition of energy in the spectrum may be checked by determining the source size for different wavelengths of light. The observed spectrum may be the overlap of several spectra from different epochs of bubble dynamics or shock waves.  In this case, it is likely that higher photon energies result from the earlier, hotter phases of the light-production mechanism, and will appear to come from a smaller radius.

Of even greater interest would be to perform the HBT experiment in those systems exhibiting discrete spectral lines from the host liquid or substances dissolved in it.  Secondary processes in the vicinity of the bubble such as the catalysis of chemical reactions, induced fluorescence or excitation by the passage of shock waves may be explored by a suitable choice of aperture size and filters. It is likely that the production mechanisms will result in interesting structure in the correlation function.  High precision measurements of the correlation function will make it possible to differentiate among various radiating structures.

The number of photons in each flash is so enormous that it is possible to 
perform most of the above measurements on an event-by-event basis.  This 
would, of course, involve the use of position-sensitive detectors with fine 
resolution. The advantage of this technique is that the flash 
characteristics are not averaged over an ensemble, hence, event by event
fluctuations can be examined directly and correlated with 
other event by event signatures of the source such as total intensity. This would be of particular relevance in the understanding of period-doubling timing
bifurcations observed for off-resonance excitation of SBSL [14].

Despite the great confidence in the spherical symmetry of the collapsing bubble, Taylor instabilities may lead to deformations of the shock front.  One possible mechanism for this could be the uneven distribution of photon energies among the normal modes of the cavity.  Since the size of the cavity is only several times the wavelength of the radiation, the excitation of non spherically-symmetric modes of the cavity would result in the nodes of the electric field dividing the surface of the sphere according to the dominant tessoral harmonic.  The resulting roughly-faceted surface could be crudely imaged using 3 or 4 photon correlations.  

\section{Conclusions}

We emphasize the ideal nature of single bubble sonoluminescence for study by means of Bose-Einstein correlations. The high brightness, reproducibility and weakness of final state interactions combine to allow us to infer for the first time, in detail, the structure of a radiating body by means of an intensity interferometer. In turn, SBSL may be a strong impetus to the development of the theory of these correlations; especially those of higher order.

  \begin{figure}[h]
    \begin{center}\mbox{ }
          {\psfig{file=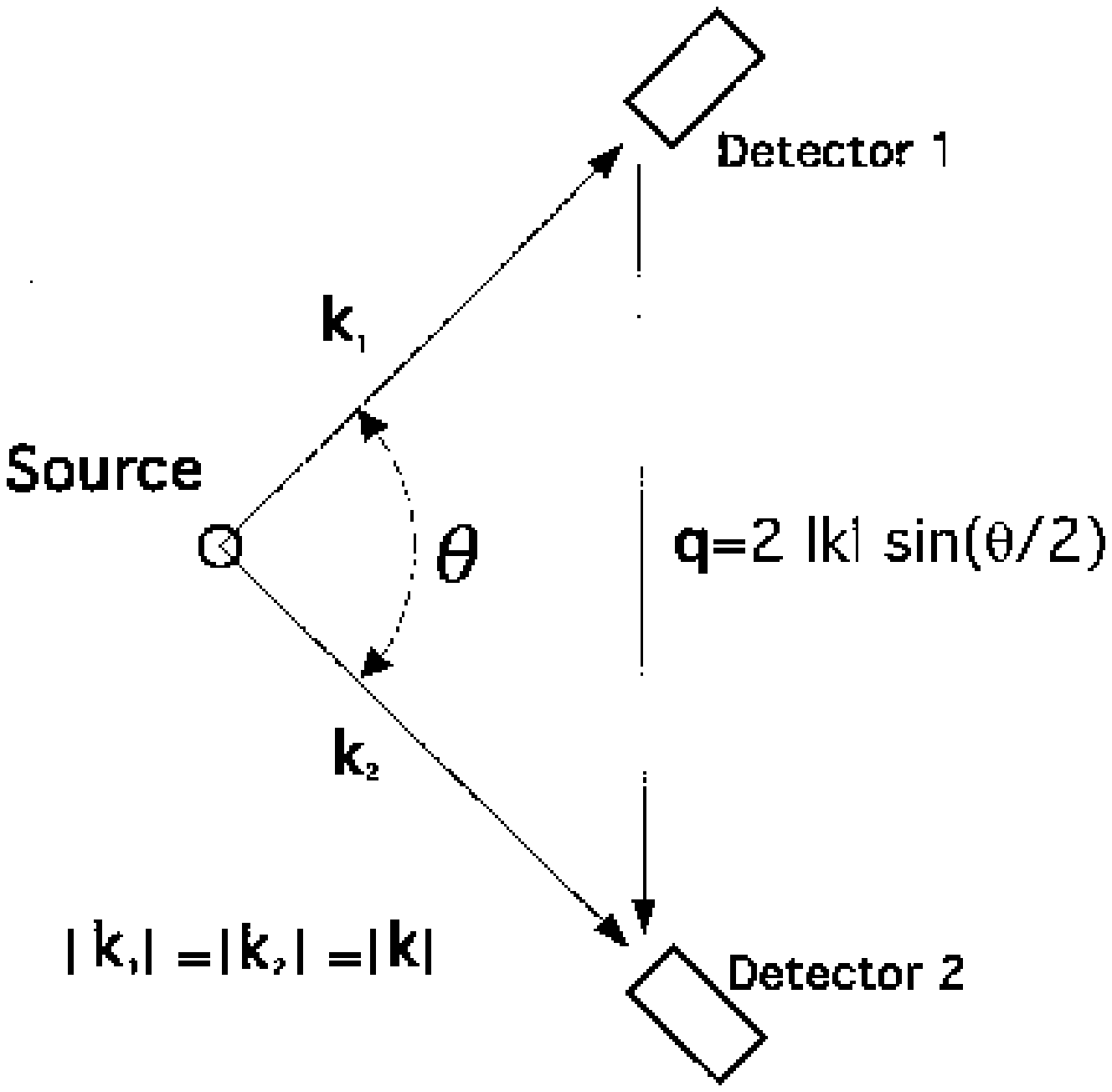,width=12.0cm}}\end{center}
    \caption{Schematic diagram of an intensity interferometer.}
    \label{fg:figure_1}
  \end{figure}
\newpage
  \begin{figure}[h]
    \begin{center}\mbox{ }
          {\psfig{file=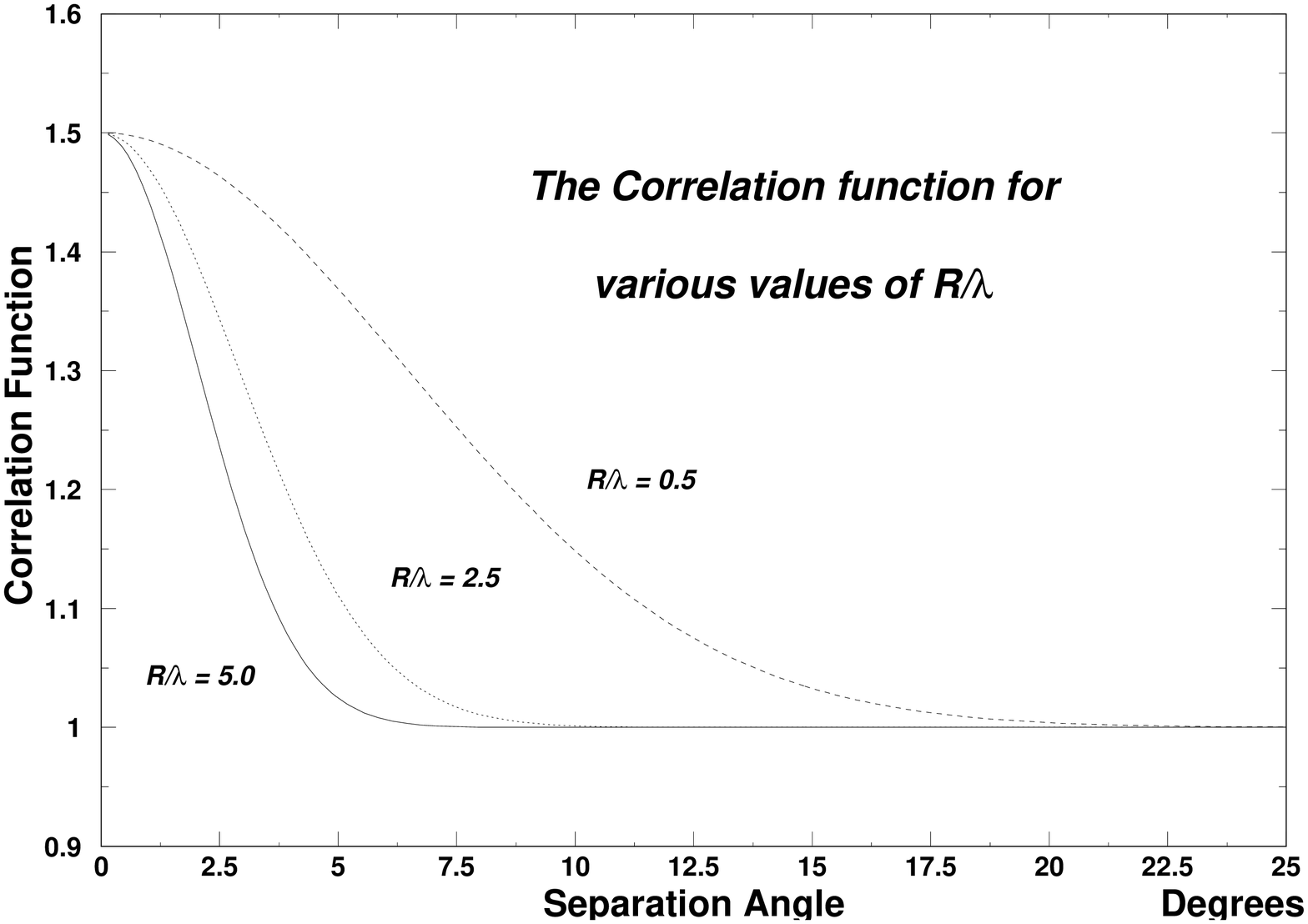,width=12.0cm}}\end{center}
    \caption{The Correlation function and its dependence on the source 
             radius and the measuring wavelength.}
    \label{fg:figure_2}
  \end{figure}


\begin{references}
 
\bibitem{RefTag}D.F. Gaitan and L.A.Crum, J. Acoust. Soc. Am Suppl. 1 {\bf 87}, S141 (1990).

\bibitem{RefTag}D.F. Gaitan, L.A. Crum, C.C. Church and R.A. Roy, J. Acoust. Soc. Am. {\bf 91}, 3166-3169 (1992).

\bibitem{RefTag}Bradley P. Barber and Seth Putterman, Nature (London) {\bf 352}, 318-320 (1991).

\bibitem{RefTag}Bradley P. Barber, R. Hiller, K. Arisaka, H. Fetterman and S. J. Putterman, J. Acoust. Soc. Am. {\bf 91}, 3061-3064 (1992).

\bibitem{RefTag}Robert Hiller, Seth J. Putterman and Bradley P. Barber, Phys. Rev. Lett. {\bf 69}, 1182-1184 (1992).

\bibitem{RefTag}B.P. Barber and S.J. Putterman, Phys. Rev. Lett.{\bf 69}, 3839-3842 (1992).

\bibitem{RefTag} David H. Boal, Claus-Konrad Gelbke and Byron K Jennings, Rev. Mod Phys. {\bf 62}, 553-602 (1990).

\bibitem{RefTag}R. Hanbury Brown and R. Q. Twiss, Phil Mag. {\bf 45}, 663-682 (1954).

\bibitem{RefTag}R. Hanbury Brown and R. Q. Twiss, Proc Roy. Soc. {\bf A243}, 291-319 (1958).

\bibitem{RefTag}M Gyulassy, S.K. Kauffmann and Lance Wilson, Phys. Rev {\bf C20}, 2267-2292 (1979).

\bibitem{RefTag}Daniel Neuhauser, Phys. Lett. {\bf 182B}, 289-292 (1986).

\bibitem{RefTag}F.B. Yano and S.E. Koonin, Phys. Lett. {\bf 78B}, 556-559 (1977).

\bibitem{RefTag}C.C. Wu and Paul H. Roberts, Phys. Rev. Lett. {\bf 70}, 3424-3427 (1993).

\bibitem{RefTag}R.G. Holt, D.F. Gaitan, A.A. Atchley, J. Holtzfuss, Phys. Rev. Lett. {\bf 72}, 1376-1379 (1994).

\end{references}
\end{document}